\documentclass[aps,prb, preprint, superscriptaddress,showpacs,showkeys,floatfix]{revtex4-1}
\usepackage{amsmath}
\usepackage{bm}
\usepackage{amssymb}
\usepackage{graphicx}
\usepackage{epstopdf}
\usepackage{epsfig}
\usepackage{color}
\usepackage{hyperref}
\usepackage{float}
\usepackage{braket}
\usepackage{lipsum}

\begin{document}

\title{Thermo-optical interactions in a dye-microcavity photon Bose-Einstein condensate}
\author{Hadiseh Alaeian
}
\affiliation{Institut f\"ur Angewandte Physik, Universit\"at Bonn, Wegelerstr. 8, D-53115 Bonn, Germany}
\author{Mira Schedensack}
\affiliation{Institut f\"ur Mathematik, Universit\"at Augsburg,
Universit\"atsstr. 14, 86159 Augsburg, Germany}
\author{Clara Bartels}
\affiliation{Institut f\"ur Angewandte Physik, Universit\"at Bonn, Wegelerstr. 8, D-53115 Bonn, Germany}
\author{Daniel Peterseim}
\affiliation{Institut f\"ur Mathematik, Universit\"at Augsburg,
Universit\"atsstr. 14, 86159 Augsburg, Germany}
\author{Martin Weitz}
\affiliation{Institut f\"ur Angewandte Physik, Universit\"at Bonn, Wegelerstr. 8, D-53115 Bonn, Germany}
\date{\today}

\begin{abstract}
Superfluidity and Bose-Einstein condensation are usually considered as two closely related phenomena. Indeed, in most macroscopic quantum systems, like liquid helium, ultracold atomic Bose gases, and exciton-polaritons, condensation and superfluidity occur in parallel. In photon Bose-Einstein condensates realized in the dye microcavity system, thermalization does not occur by direct interaction of the condensate particles  as in the above described systems, i.e. photon-photon interactions, but by absorption and re-emission processes on the dye molecules, which act as a heat reservoir. Currently, there is no experimental evidence for superfluidity in the dye microcavity system, though effective photon interactions have been observed from thermo-optic effects in the dye medium. In this work, we theoretically investigate the implications of effective thermo-optic photon interactions, a temporally delayed and spatially non-local effect, on the photon condensate, and derive the resulting Bogoliubov excitation spectrum. The calculations suggest a linear photon dispersion at low momenta, fulfilling the Landau's criterion of superfluidity . We envision that the temporally delayed and long-range nature of the thermo-optic photon interaction offer perspectives for novel quantum fluid phenomena. 

\end{abstract}

\maketitle

\section{Introduction}
For a gas sufficiently cold and dense that the thermal de Broglie wavelength exceeds the interparticle spacing, quantum statistical effects come into play. Specifically, for massive bosonic particles, Bose-Einstein condensation into a macroscopically populated ground state minimizes the free energy, as has been experimentally demonstrated in the gaseous regime with ultracold atoms more than 20 years ago~\cite{Ketterle02}.

More recently, Bose-Einstein condensation has also been observed with exciton-polaritons, mixed states of matter and light, and with photons~\cite{Carusotto13, Klaers101, Klaers11, Marelic15}. Unlike particles with a non-vanishing rest mass, photons usually do not show Bose-Einstein condensation. The thermal photons of blackbody radiation have no chemical potential, corresponding to a non-conserved particle number upon temperature variation. Therefore, blackbody radiation photons vanish at low temperature instead of showing a phase transition to a condensate. This difficulty was overcome in 2010 by confining photons in a dye-solution filled optical resonator made of two mirrors spaced in the micrometer regime~\cite{Klaers101, Marelic15}. The short mirror spacing effectively imprints a low-frequency cutoff for the photon gas, and in the presence of a mirror curvature the problem becomes formally equivalent to a two-dimensional system of harmonically confined massive bosons. By repeated absorption and re-emission processes, photons thermalize to the (rovibrational) temperature of the dye solution at room temperature.

Experimentally, both the thermalization of the photon gas to the dye temperature~\cite{Klaers10} and Bose-Einstein condensation of photons above a critical particle number have been observed~\cite{Klaers101, Marelic15}. For larger condensate fractions, the size of the condensate increases, which is attributed to a weak repulsive interaction mainly due to the thermo-optic effects. 

So far, it is not known if the photon gas, in addition to exhibiting Bose-Einstein condensation, is a superfluid~\cite{Snoke13}.
The existence of superfluidity is believed to require direct interparticle interactions, as e.g. present for polaritons, for which superfluidity has been established~\cite{Carusotto04, Amo09, Nardin11, Sanvitto11, Amo11, Grosso11}. On the theoretical side, the concept of a nonlinear photon fluid was first introduced by Brambilla et al. and Staliunas~\cite{Brambilla91, Staliunas93}, who used hydrodynamic equations to describe electromagnetic fields in a cavity. Chiao et al. subsequently proposed to generate a photon superfluid in a nonlinear optical cavity using the Kerr-effect and furthermore predicted sound-like modes at the low-momentum part of the Bogoliubov dispersion~\cite{Chiao99, Chiao00}. 

In this paper, we examine the effect of thermo-optic interactions on the dispersion of a photon gas trapped inside an optical microcavity. The thermo-optic effect, also known as thermal lensing, was first introduced by J. Gordon while studying transient effects of the output power and the beam size upon inserting a liquid cell inside the laser cavity~\cite{Gordon65}. In a propagating configuration in stationary conditions, the temporal delay of the thermo-optic effect does not play a role. The signatures of superfluidity and non-local effects from the associated thermo-optic nonlinearity have been experimentally observed in this configuration~\cite{VOCKE15,Vocke16}. In another work, Strinati and Conti have theoretically studied the stationary state of dye microcavity photon condensate subject to a non-local thermo-optic nonlinearity~\cite{Strinati14}.

We report a theoretical study on the effects of a thermo-optic nonlinearity giving rise to a temporally delayed and non-local effective photon interaction for the photon condensate in the dye-microcavity system. Assuming a plane mirror microcavity geometry, we derive the Bogoliugov dispersion for such a system subject to small perturbations. For a suitable parameter range, we find a linear dispersion at low momentum, corresponding to a phonon-type dispersion. We discuss the possibility of superfluidity of the photon condensate based on such temporally delayed and long-range interactions.

In the following, Chapter II discusses some general properties of the photon gas subject to the thermo-optic interactions, and section III gives steady-state solution. In section IV we derive Bogoliubov modes of the system and the resulting elementary excitation dispersion. Finally, section V concludes the manuscript.

\section{Thermo-optic interactions in a photon gas}
\begin{figure*}[h]
\centering
\includegraphics[scale=0.9]{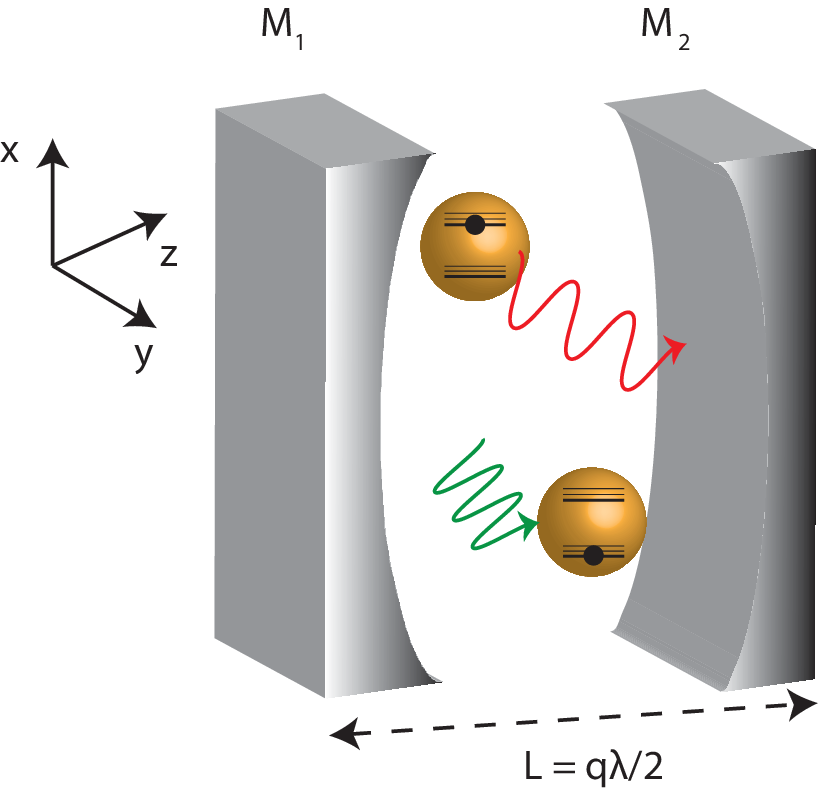}
\caption{\label{Fig1} Schematics of the dye-filled microcavity setup. The mirrors $M_{1,2}$ are spherically curved with radius $R$, leading to a harmonic trapping potential for the photon gas. The optical cavity imposes a non-zero cut-off frequency for the trapped photons, with the longitudinal modal quantum number $q$. The remaining transverse modal quantum numbers make the photon gas two-dimensional. The photon gas thermalizes by repeated absorption re-emission processes to the spectral temperature of the dye molecules.}
\end{figure*}

We begin by discussing some general formulas describing the system of a photon gas trapped in a dye microcavity subject to thermo-optic interactions. In the experimental scheme of~\cite{Klaers101, Klaers11}, Bose-Einstein condensation of photons is achieved in a dye microcavity (see Fig.~\ref{Fig1}) with the mirror spacing in the wavelength regime. This leads to a large frequency spacing between longitudinal modes, comparable to the emission width of the dye molecules. Therefore, to good approximation only photons of a fixed longitudinal mode order $q$ are found in the resonator, and the two remaining transverse mode quantum numbers make the system two-dimensional. The transverse $TEM_{00}$-mode has the lowest allowed frequency, which imposes a low-frequency cutoff. Moreover, the photon dispersion becomes quadratic, i.e. massive particle-like, and the mirror curvature imposes a harmonic confinement on the photon gas. 

One can show that the photon gas in the cavity is formally equivalent to a two-dimensional gas of massive bosons with effective mass $m_{ph}=\hbar k_zn_0/c = \hbar\omega_c (n_0/c)^2$ , where $\omega_c$ denotes the cutoff frequency, $k_z$ the longitudinal wavevector, and $n_0$ the refractive index of the solution. The photon energy in the paraxial limit is:

\begin{equation}
\label{photon energy}
E_{ph}\simeq m_{ph}c^2 + \frac{(\hbar k_r)^2}{2m_{ph}}+V_{trap}(x,y)+E_{int}~,
\end{equation}

where $k_r$ denotes the transverse momentum and $V_{trap} (x,y)= \frac{1}{2}m_{ph}\Omega^2(x^2 + y^2)$ is the trapping potential. The trapping frequency of the harmonic potential is given by $\Omega = (c/n_0)/\sqrt{LR/2}$, where $L$ and $R$ are the mirror spacing and curvature, respectively. Finally $E_{int}$ is the effective photon interaction energy, as will be discussed below.

Thermal equilibrium of the photon gas is achieved by repeated absorption and re-emission processes by the dye molecules. For the described two-dimensional, harmonically confined system it is known that a Bose-Einstein condensate exists at finite temperature. Accounting for the two-fold polarization degeneracy of photons, one finds the critical particle number: 

 \begin{equation}
 \label{photon critical number}
 N_c = \frac{\pi^2}{3}(\frac{k_BT}{\hbar\Omega})^2~.
 \end{equation}
 
For $\Omega = 2\pi \times 3.6\times 10^{10}~Hz$, as derived for $L=2~\mu m$, $R=1~m$, and $n_0=1.33$, at room temperature ($T=300^o~K$) one obtains $N_c\approx99000$ for the critical particle number.

During the course of the absorption re-emission processes of the dye molecules, a small fraction of inelastic processes due to dye's finite quantum efficiency ($\eta\simeq 95\%$ for the case of rhodamine dye solution) causes local heating of the solvent. Due to the temperature dependence of the solution refractive index, the optical distance between the mirrors is decreased at the corresponding transverse position in the cavity. This is equivalent to a local rise of the photon gas potential. In other words, the heating with a corresponding decrease of the refractive index results in a smaller optical wavelength, hence a higher photon energy is required to locally match the mirror boundary conditions.  The resulting interaction energy is:

\begin{equation}
\label{interaction energy}
E_{int}\simeq -m_{ph}(\frac{c}{n_0})^2\frac{\Delta n}{n_0}~,
\end{equation}
 
where $\Delta n =\beta \Delta T$ , with $\beta = \partial n/\partial T$ as the thermo-optic coefficient of the solution. 

The photon gas is well described by an equilibrium state, if thermalization by coupling to the dye molecules is faster than both loss and pump processes. In this case, photons can relax towards the ground state of the harmonic trapping potential and form a Bose-Einstein condensate before they are lost through mirror transmission or inelastic processes in the dye~\cite{Kirton13, Schmitt15}. Accounting for the thermo-optic effective photon interactions, the condensate dynamics can be described with a generalized time-dependent Gross-Pitaevskii equation:

\begin{equation}
\label{generalized GPE}
i\hbar\frac{\partial\psi(\vec{r},t)}{\partial t}=\Bigg(-\frac{\hbar^2\nabla^2}{2m_{ph}} - m_{ph}(\frac{c}{n_0})^2\frac{\beta}{n_0}\Delta T(\vec{r},t)\Bigg) \psi(\vec{r},t)~.
\end{equation}

In the above equation $\psi(\vec{r},t)$ denotes the slowly varying envelope of the condensate wavefunction (in the mean-field treatment) in the rotated frame of $\omega_c$. Note that the above equation is more general and captures the full 3D behavior of the wavefunction as ($\psi(x,y,z,t)$). Also instead of using the effective trapping potential of eq.~\ref{photon energy}, we impose Dirichlet boundary conditions on the spherical mirror surfaces. Similar as in cold atom physics literature~\cite{Brachet12,Salazar13}, here we do not consider interaction effects of the thermal cloud, due to its much lower density.

The time evolution of the relative local temperature is determined by the heat transport equation as:

\begin{equation}
\label{heat transport equation}
\frac{\partial \Delta T(\vec{r},t)}{\partial t} = \frac{\kappa}{C_v}\nabla^2\Delta T(\vec{r},t) + \frac{\alpha_{in} c \hbar\omega}{C_v n_0}|\psi(\vec{r},t)|^2~,
\end{equation}

The first term on the right hand side accounts for the heat diffusion and the second term for heating through inelastic processes in the dye solution. The parameters $\kappa$ and $C_v$ are the thermal conductivity and volume heat capacity, respectively. Further, $\alpha_{in}$ is the inelastic absorption coefficient and is related to the absorption coefficient $\alpha$ via the quantum efficiency of the dye $\eta$, as $\alpha_{in} = (1-\eta)\alpha$.

Moreover, the following normalization condition for the wavefunction is held:
 
\begin{equation}
\label{WF normalization condition}
\int_V dv|\psi(\vec{r})|^2= N_{BEC}~,
\end{equation}

where $N_{BEC}$ is the total number of photons in the condensate. Table~\ref{Table1} gives relevant parameters of a typical dye-filled cavity setup with methanol solvent~\cite{Schmitt15}, \cite{Lusty87}.

\begin{table}
\label{parameter table}
\begin{center}
\begin{tabular}{|c|c|c|c|c|c|c|c|c|c|c|}
\hline
$L (\mu m$) & $\alpha_{in} (m^{-1})$ & $n_0$ & $\beta=\frac{dn}{d\tilde{T}} (K^{-1})$  & $C_v (J K^{-1} m^{-3})$ & $\kappa (W m^{-1} K^{-1})$ \\
\hline
2 & 0.63  & 1.33 & -4.8$\times 10^{-4}$ & 1.9$\times 10^6$ & 0.168 \\
\hline
\end{tabular}
\caption{\label{Table1} List of physical parameters of a dye-filled cavity setup used in this paper. The properties of the solvent, methanol, are from.~\cite{Lusty87} The value of the inelastic absorption coefficient is for 1mM solution of R6G in methanol solvent and is calculated from the experimental data in~\citep{Schmitt15}, assuming a quantum yield of $95\%$ for the dye~\cite{Bindhu99}.}
\end{center}
\end{table}

\section{Steady-State}
In the steady state, the temperature is settled to a stationary value $T_{ss}(\vec{r})$, and the the time evolution of the condensate wavefunction is given by $\psi(\vec{r};t)=\psi_{ss}(\vec{r})e^{-i\mu t}$. The stationary forms of eq.~\ref{generalized GPE} and \ref{heat transport equation} are:

\begin{subequations}
\label{steady state equations}
\begin{eqnarray}
\hbar\mu\psi_{ss}(\vec{r})=\Bigg(-\frac{\hbar^2\nabla^2}{2m_{ph}}-m_{ph}(\frac{c}{n_0})^2\frac{\beta}{n_0} T_{ss}(\vec{r})\Bigg) \psi_{ss}(\vec{r})~,\\
\nabla^2T_{ss}(\vec{r})+\frac{\alpha_{in} c \hbar\omega}{n_0\kappa}|\psi_{ss}(\vec{r})|^2=0~.
\end{eqnarray}
\end{subequations}

The mirrors ($M_{1,2}$ in Fig.~\ref{Fig1}) are macroscopically large. Moreover, their thermal conductivity exceeds that of the dye solution, so in all of the calculations we assume $\Delta T=0$ on the mirror surfaces.

\subsection{Numerical solution}
In the presence of a mirror curvature, as required to obtain a trapping potential, the coupled sets of eq.~\ref{steady state equations} can only be solved numerically. Without a thermo-optic non-linearity (i.e. for $\beta=0$), the ground state of the harmonic trap is the lowest energy state. Following the usual convention we normalize the chemical potential to yield $\mu = 0$ in this interaction-less case. 
For earlier work investigating the steady state properties of a thermo-optic interaction in the harmonically trapped condensate inside a microcavity, see Ref.~\cite{Strinati14}.

\begin{figure*}[h]
\centering
\includegraphics[scale=0.85]{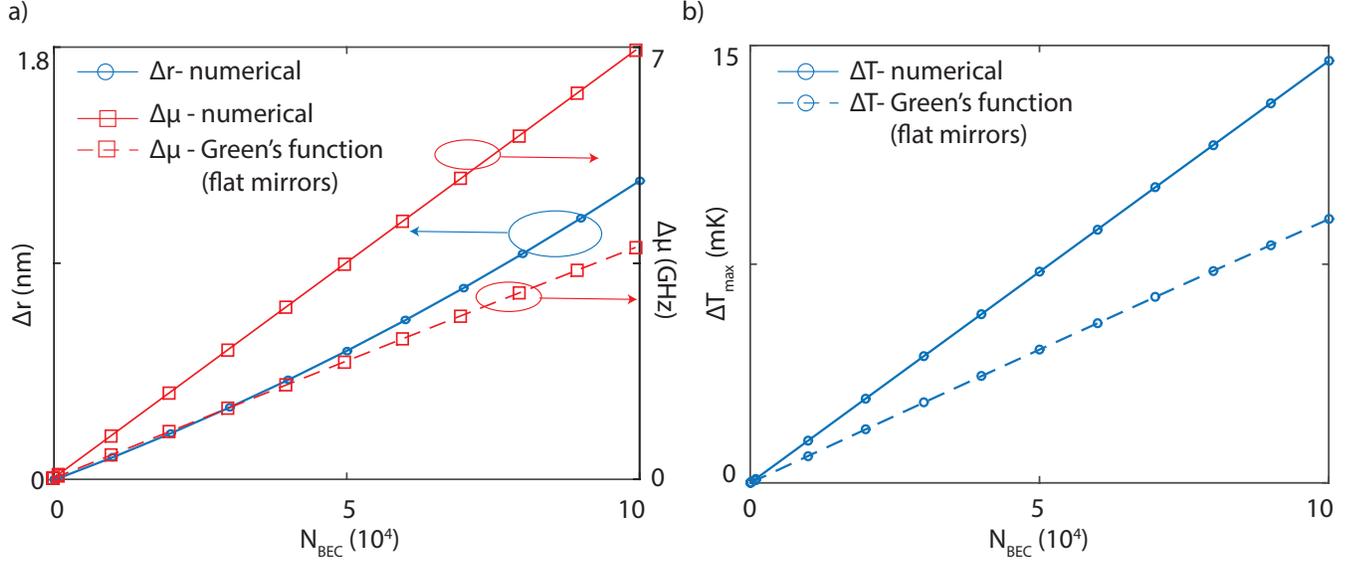}
\caption{\label{Fig2} (a) Variation of the condensate radius (solid blue line with circles) and the chemical potential (solid red line with squares) as a function of number of photons in the condensate, as obtained by numerical solution of the problem for cavity mirrors of radius $R=1~m$.
For comparison, the red dashed line with squares gives the  variation of the chemical potential obtained using the Green's function method for the problem with flat mirrors. (b) The solid blue line with circles gives the maximum temperature increase in the dye microcavity as the number of photons in the condensate, obtained from the numerical method for curved mirrors with $R=1~m$ radius. The dashed blue line with circles shows the corresponding results obtained using the Green's function approach for flat mirrors. The results are for $L = 2~\mu m$ mirror spacing and absorption coefficient $\alpha_{in} = 0.63~m^{-1}$. For the Green's function method calculation investigating the flat mirror problem, the quoted value for the photon number refers to the number of photons in an area $\pi r_{BEC}^2$. $r_{BEC}$ corresponds to the interaction less condensate radius of the curved mirror problem. In this way, the assumed average photon density is the same for both of the numerical and the Green's function approaches.}
\end{figure*}

Using a fully numerical algorithm we solved the above coupled non-linear equations to investigate the effect of non-linearity on the interacting condensate. The results for a symmetric cavity with mirrors of $R=1~m$ radius of curvature are shown by the solid lines in Fig.~\ref{Fig2}. In the absence of the thermo-optic interaction the Gaussian condensate radius is $r_{BEC} = \sqrt{1 \hbar/2m_{ph}\Omega} \simeq 6~\mu m$ for the quoted values of the cavity length and mirror radius of curvature. 

Panel (a) shows the variation of the condensate radius from the interaction-less case $\Delta r$ and the chemical potential $\Delta \mu$ as a function of number of photons in the condensate. As can be seen both of these parameters linearly increase as the photon number becomes larger, implying a larger condensate with a higher energy. This is consistent with the physical consequence of a repulsive interaction mediated by thermo-optic non-linearity. 
To clarify the behavior further, Fig.~\ref{Fig2}(b) shows the maximum value of the temperature increase in the dye microcavity as a function of condensate photon number $N_{BEC}$. As can be seen the temperature monotonically increases with increasing number of photons, giving rise to a larger change of the refractive index, hence a stronger effective photon interaction.

A comparison with experimental results is not straightforward since the condensate mode diameter increase reported in~\cite{Klaers10} corresponds to the accumulative stationary value observed for a pulsed pump with certain repetition rate. The values here are nevertheless smaller than the experimentally observed ones. We attribute this discrepancy to the boundary conditions used for the temperature distribution, which imposes that $\Delta T$ vanishes at the mirror surfaces. A more realistic model in the presence of the thermo-optic non-linearity needs to account for the finite thermal-conductivity of the mirrors and include the thermal properties of both mirror layers as well as the substrate. However, we expect the dynamic properties of the condensate, as discussed in the following, to be less affected by the thermal properties of the mirrors since the local properties of the heat transfer will dominate in that case.

\subsection{Green's function approach}
Aside from being computationally expensive, the fully numerical method fails to provide one with more physical insight about the problem. In this section we propose an alternative method by employing the Green's function of the heat diffusion problem, making the analysis more efficient. Moreover, this form could be used further for elementary excitation studies as will be discussed in the next sections.

The coupled eigenvalue problem of eq.~\ref{steady state equations} can be efficiently solved using Green's function of heat diffusion problem when proper boundary conditions are applied. The Green's function of the heat transfer problem will be determined as:

\begin{subequations}
\label{Green's function of heat}
\begin{eqnarray}
\nabla^2G_{NL}(\vec{r};\vec{r}')=\delta(\vec{r}-\vec{r}')~,\\
G_{NL}(\vec{\rho}, z=\pm L/2;\vec{r}')=0~.
\end{eqnarray}
\end{subequations}

For simplicity in our analytic calculations the mirrors are assumed to be flat at $z=\pm L/2$, and extended to infinity in the transverse plane ($xy$-plane).

As eq.~\ref{Green's function of heat} are isomorphic to an electrostatic problem, image theory can be used to find a closed-form Green's function. Using proper positive and negative images at $(\pm 2nL+z')$ and $(\pm (2n+1)L-z')$ respectively, we obtain: 

\begin{equation}
\label{Green's function closed form}
G_{NL}(\vec{r};\vec{r}')=-\frac{1}{4\pi}\sum_{n=-\infty}^{+\infty}(\frac{1}{\sqrt{|\vec{\rho}-\vec{\rho}'|^2+(z-z'+2nL)^2}}
-\frac{1}{\sqrt{|\vec{\rho}-\vec{\rho}'|^2+(z+z'+(2n+1)L)^2}})~,
\end{equation}

where $\rho=\sqrt{x^2+y^2}$ and $\rho'=\sqrt{x'^2+y'^2}$.

When combined with the eigenvalue problem of the wavefunction in eq.~\ref{steady state equations} one obtains the following equation for the steady-state:

\begin{equation}
\label{GPE-steady state}
\hbar\mu\psi_{ss}(\vec{r})=\Bigg(-\frac{\hbar^2\nabla^2}{2m_{ph}}+m_{ph}(\frac{c}{n_0})^3\frac{\beta\alpha_{in} \hbar \omega}{n_0\kappa} \int_{V_c} dv G_{NL}(\vec{r};\vec{r}')|\psi_{ss}(\vec{r}')|^2\Bigg) \psi_{ss}(\vec{r})~.
\end{equation}

Equation~\ref{GPE-steady state} has the form of a Gross-Pitaevskii equation, while the interaction term is different from the usual contact form. Here, the interaction potential has an integral form describing a non-local interaction given by the Green's function, where the strength of the interaction potential at point $\vec{r}$ is related to the whole condensate wavefunction distribution.

\begin{figure*}[h]
\centering
\includegraphics[scale=0.8]{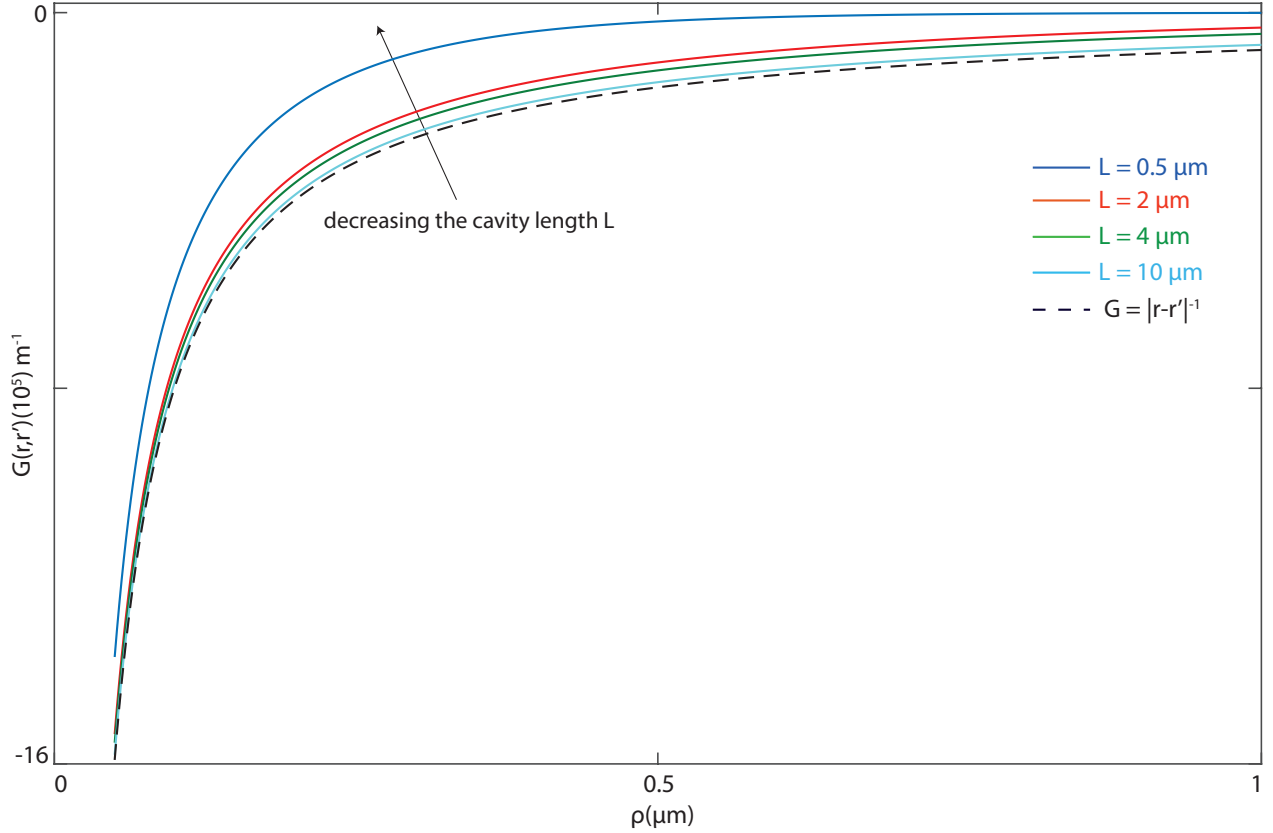}
\caption{\label{Fig3} Variation of the Green's function $G_{NL}(\vec{r},\vec{r}')$ vs. $\rho$ (the radial parameter) in the transverse plane when $\vec{r}'=0, z=0$. The effect of mirror separation and non-locality are compared for different $L$. As can be seen the Green's function extends further with an increases in the cavity length. This means that the interaction range is longer in thicker cavities. For large mirror spacing, the Green's function is in good agreement with the gravitational-type dependency given by $1/|\vec{r}-\vec{r}'|$ (black dashed line).}
\end{figure*}

Figure~\ref{Fig3} shows the variation of the Green's function $G_{NL}(\vec{r};0)$ as a function of $\rho$ in the transverse plane for different mirror separations. As can be seen at a fixed radial point the value of the Green's function increases as the cavity length becomes larger.  This behavior implies that the temperature changes from a 2D distribution to 3D in thicker cavities. In the limit of very large mirror spacing the Green's function approaches a Coulomb-type long-range interaction given by $G_{NL}(\vec{r};\vec{r}')\approx1/|\vec{r}-\vec{r}'|$. This behavior can be clearly observed in Fig.~\ref{Fig3} where the thick cavity Green's function ($L=10~\mu m$) and the asymptotic Columbic form (black dashed line) are in a good agreement. 

To compare the predictions of these two approaches, we have used the generalized Gross-Pitaevskii eq.~\ref{GPE-steady state} to calculate the stationary features of the condensate. Unlike the numerical method however, the Green's function approach only allows for a treatment of a flat mirror geometry. The red dashed line with squares in Fig.~\ref{Fig2}(a) shows the  chemical potential variation for the homogeneous problem calculated with this method. Similarly, the dashed line with circles in Fig.~\ref{Fig2}(b) shows the corresponding variation of the maximum temperature with the number of photons in the condensate. 
The obtained results from the Green's function approach are in approximate agreement with the numerical results for the problem with curved mirrors, and show the same trend. We point out that an exact agreement is not expected here due to the different mirror geometries, with correspondingly different condensate mode profiles.

In contrary to the case of atomic BECs where the range of interaction is strictly limited by the interaction type, the photon fluid shows a unique feature of possessing a tunable interaction range from a local form to a highly non-local gravitational type interaction in relatively thin and thick cavities, respectively.
In practice, the requirement of photon Bose-Einstein condensation for a low-frequency cutoff imposes a maximum usable cavity length for corresponding experiments.

\section{Small perturbations and Bogoliubov modes}
For weak perturbations the dynamics of the system can be found by assuming small fluctuations around the stationary solutions. We use the following ansatz to determine the Bogoliubov modes, where $\Omega$ denotes the frequency of the perturbations.

\begin{subequations}
\label{small perturbation 1}
\begin{align}
\psi(\vec{r};t)=(\psi_{ss}(\vec{r})+u(\vec{r})e^{-i\Omega t}+v^*(\vec{r})e^{+i\Omega^* t})e^{-i\mu t}~,\\
T(\vec{r};t)=T_{ss}(\vec{r})+\delta t(\vec{r})e^{-i\Omega t}+\delta t^*(\vec{r})e^{+i\Omega^* t}~.
\end{align}
\end{subequations}

Inserting this ansatz into the original non-linear equations and neglecting the terms with orders higher than one in the perturbation, one can derive the following linear coupled equations for the small fluctuations $u(\vec{r}), v(\vec{r})$, and $\delta t(\vec{r})$:

\begin{equation}
\label{small perturbation 2}
\Omega\begin{bmatrix}
u(\vec{r})\\ v(\vec{r})\\ \delta t(\vec{r})
\end{bmatrix}
=
\mathcal{L}
\begin{bmatrix}
u(\vec{r})\\ v(\vec{r})\\ \delta t(\vec{r})
\end{bmatrix}~,
\end{equation}

where:
\begin{equation}
\label{Bogoliubov matrix}
\mathcal{L}=
\begin{bmatrix}
[(-\frac{\omega}{2}-\mu)-\frac{\omega}{n_0}\beta T_{ss}-\frac{c^2}{2\omega n_0^2}\nabla^2] & 0 & -\frac{\omega}{n_0}\beta\psi_{ss} \\
0 & -[(-\frac{\omega}{2}-\mu)-\frac{\omega}{n_0}\beta T_{ss}-\frac{c^2}{2\omega n_0^2}\nabla^2] &  \frac{\omega}{n_0}\beta\psi^*_{ss}\\
i\frac{\alpha n_0}{2C_vZ_0}\psi^*_{ss} & i\frac{\alpha n_0}{2C_vZ_0}\psi_{ss} & i\frac{\kappa}{C_v}\nabla^2\\
\end{bmatrix}~.
\end{equation}

To begin with, we derive a solution that neglects the temporal delay of the temperature and assumes that it follows the time dependency of the condensate density $|\psi(\vec{r},t)|^2$. With this assumption the temperature change $\Delta T(\vec{r},t)$ is given by a diffusion-type equation, as for $T_{ss}$ in eq.~\ref{steady state equations}, which yields a non-local, but instantaneous effective interaction. Later in this section we will modify this assumption to take into account the temporal delay of the temperature distribution due to the finite heat conductivity.

For a translationally invariant problem, a plane wave ansatz of the form $e^{i\vec{k}\cdot\vec{r}}$ would be well suited to describe the spatial part of the excitations. More suited for this problem, given that the condensate wavefunction $\psi(\vec{r})$ is not spatially uniform, is the Green's function approach, building upon the stationary eigenfunctions discussed in the previous section. The mirrors impose Dirichlet boundary conditions and break the translational invariance along the $z$-axis. Therefore, we approximate the longitudinal variation of the wavefunction as $\sqrt{2/L}~ sin(q\pi z/L)$, and define an effective 2D Green's function in the transverse plane:

\begin{equation}
\label{2D GF}
G_{eff}^{2D}(\vec{\rho},\vec{\rho}') =  \int dz dz' sin^2(\frac{q\pi z}{L}) sin^2(\frac{q\pi z'}{L}) G_{NL}^{3D}(\vec{r},\vec{r'}) 
=-\frac{1}{\pi L}\sum_{n_{odd}} (\frac{8q^2}{n\pi(n^2-4q^2)})^2
K_0(\frac{n\pi}{L}|\vec{\rho}-\vec{\rho}'|)~,\
\end{equation}

where $K_0$ is the $0^{th}$-order modified Bessel function.

As can be inferred, this effective 2D potential is translationally invariant in the transverse plane, and leads to a well-defined transverse momentum $\vec{k}$ for the elementary excitation. Therefore, the Bogoliubov dispersion of this system is properly defined for frequency $\Omega$ and the transverse momentum $\vec{k}$, and from eq.~\ref{small perturbation 2} is determined as:

\begin{equation}
\label{non-local dispersion1}
\Omega^2=\frac{k^2}{2m_{ph}}\bigg(\frac{\hbar^2k^2}{2m_{ph}} + (2\hbar\omega)^2|\psi_{ss}(0)|^2\frac{\alpha\beta c}{n_0^2\kappa}\hat{G}_{NL}(k)\bigg)~,\\
\end{equation}
with
\begin{equation}
\label{non-local dispersion2}
\hat{G}_{NL}(k) = -\frac{2}{L}\sum_{n_{odd}} (\frac{8q^2}{n\pi (n^2-4q^2)})^2
(\frac{L}{n\pi})^2 \frac{1}{1+(\frac{L}{n\pi})^2k^2}~.
\end{equation}

Here $\hat{G}_{NL}(k)$ is the Fourier transform of $G^{2D}_{eff}(\vec{\rho})$, and in the first formula $|\psi_{ss}(0)|^2$ can be approximated as:

\begin{equation}
\label{psi_ss0}
|\psi_{ss}(0)|^2\approx \frac{N_{BEC}}{\pi {r_{BEC}}^2L}~.
\end{equation}

Equation~\ref{non-local dispersion1} gives the dispersion in the presence of the thermo-optic interaction. At large momenta, the first term on the right hand side dominates, yielding the usual particle-like quadratic dispersion of photons, see also eq.~\ref{photon energy}. In other words, when the momentum is large the interactions have negligible effect. 

The thermo-optic interactions impact the low-momentum part of the dispersion, and the corresponding effect is predominantly determined by the function $\hat{G}_{NL}(k)$, i.e. the Fourier transform of the Green's function. For wavevectors $k\le k_z = q\pi/L$, within the range that the paraxial limit is fulfilled, $\hat{G}_{NL}(k)$ becomes almost $k$-independent, hence a linear tendency for low-momentum excitations is expected.

So far we have ignored the explicit dynamics of the delayed temperature given by the $\dot{T}$-term in the heat equation. This effect can be taken into account by employing the proper Green's function of the heat equation in eq.~\ref{heat transport equation} which depends on time as well as the position. This time-dependent Greens' function modifies the aforementioned dispersion of eq.~\ref{non-local dispersion1} to a transcendental equation for the dispersion (i.e. $\Omega(k)$) after substituting $\hat{G}_{NL}(k)$ with $\hat{G}_{NLD}(\Omega,k)$ as:

\begin{equation}
\label{delayed, non-local dispersion}
\hat{G}_{NLD}(\Omega,k) = -\frac{2}{L}\sum_{n_{odd}} (\frac{8q^2}{n\pi (n^2-4q^2)})^2\times
\frac{1}{(\frac{n\pi}{L})^2-i\Omega} \frac{1}{1+k^2\frac{1}{(\frac{n\pi}{L})^2-i\Omega}}
\end{equation}

Equation~\ref{delayed, non-local dispersion} together with eq.~\ref{non-local dispersion1} gives the final results for the quasi-particle dispersion, including the effects of both non-locality and the delay in such thermo-optic interactions. As the form of the transcendental equation suggests, the delayed nature of the interaction leads to complex frequencies $\Omega$, implying an instability of the condensate. Compared to the non-local case only, this is the main qualitative modification of the temporal effect. We notice this behavior can be compared with the dynamical instability predicted and observed in polariton condensates~\citep{Wouters07}. While the instability of the latter is due to the coupling of the condensate with the exciton reservoir, in this dye microcavity system the condensate instability occurs due to the thermal coupling to the dye solutions. At longer times, we expect that the thermo-optic interaction destroys the photon condensate for larger interaction strengths.

\begin{figure*}[h]
\centering
\includegraphics[scale=0.85]{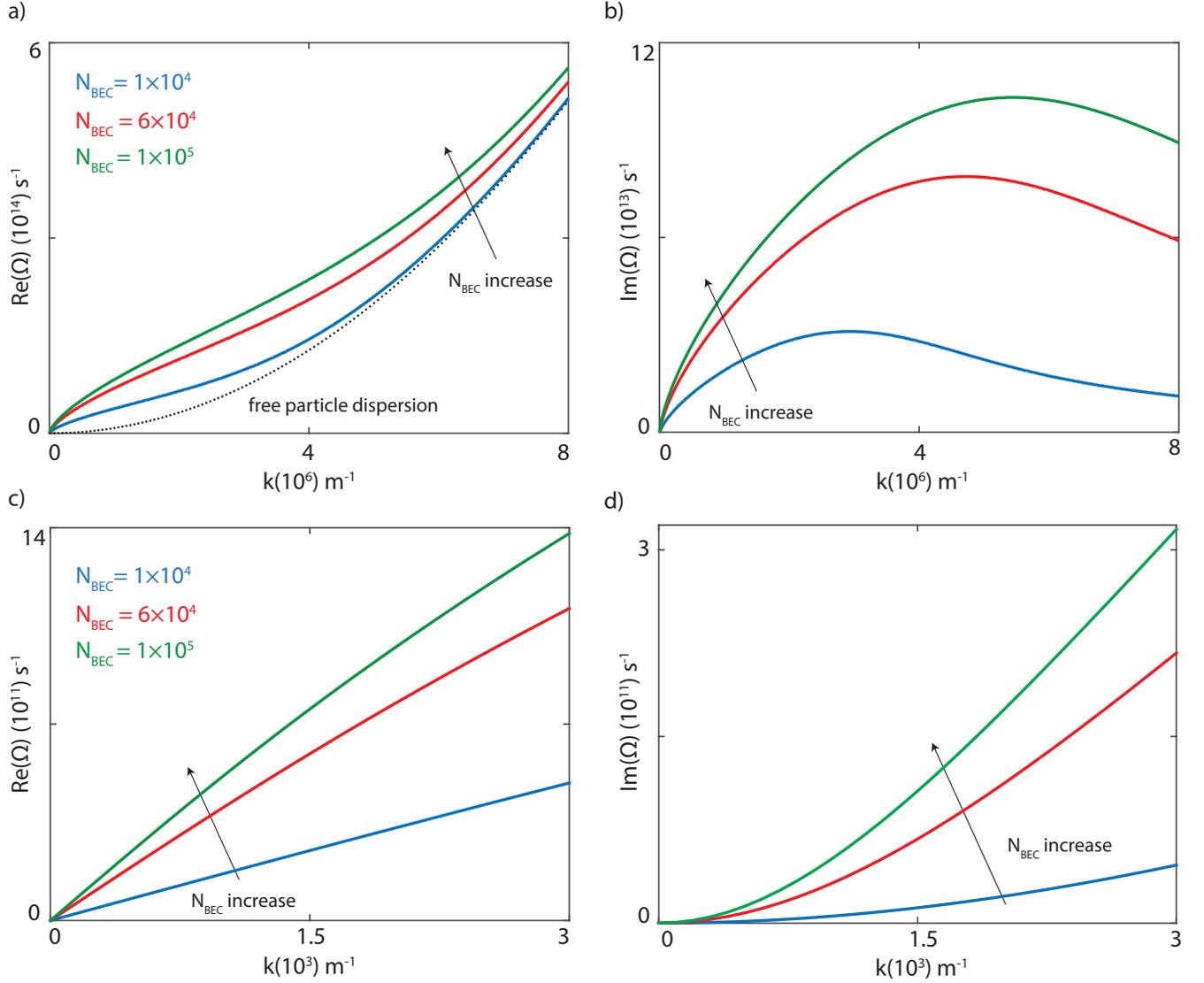}
\caption{\label{Fig4} (a) Real and (b) imaginary part of the quasi-particle dispersion in the presence of a thermo-optic effective photon interaction, given by eq.~\ref{non-local dispersion1} and eq.~\ref{delayed, non-local dispersion} for different photon numbers $N_{BEC}$. Other parameters are  $L=2~\mu m$, $\alpha_{in} = 0.63~ m^{-1}$, and $q=9$. In (a) the black dotted line shows the free-particle dispersion for comparison. The effect of interaction in modifying the low-momentum part of dispersion from this free-particle tendency is increased as more photons are put in the condensate. At larger momenta all condensates seems to behave as free-particles which their dispersion tends to behave quadratically as free-particles (a), with zero imaginary parts (b).
Zoomed-in (c) real and (d) imaginary part of the dispersion at very low momentum. For this range the real part behaves linearly and the imaginary part is small, suggesting a superfluid behavior.}
\end{figure*}

Figures~\ref{Fig4}(a),(b) show the real and imaginary part of the dispersion respectively for various number of photons in the condensate $N_{BEC}$. The thick, dotted black line in panel (a) gives the (quadratic) free-particle dispersion in the absence of interactions. As stated above, at large momenta this free particle behavior is approached, and the imaginary part asymptotically approaches zero (Fig.~\ref{Fig4}(b)), leading to a stable condensate with a quadratic dispersion~\footnote{For a momentum $k$, both $\Omega$ and $-\Omega^*$ are solutions of the dispersion equations. However as the behavior of these two branches are not different here we only show the results for the solution with positive real part.}.

The low-momentum behavior however, noticeably deviates from the non-interacting dispersion. The difference between the interacting photon fluid dispersion and the ideal one increases for larger photon numbers. Figure~\ref{Fig4}(c) and (d) show the zoomed-in real and imaginary part of Bogoliubov dispersion at very low momenta. The linear dispersion behavior of Fig.~\ref{Fig4}(c), accompanied with low imaginary value as in Fig.~\ref{Fig4}(d), means that the low momentum quasi-particles behave like phonons, and move with a constant velocity $v_c$ in the photon condensate. This feature suggests that the photon BEC can potentially be a superfluid, a feature that can be better understood using the non-local effect of the Green's function.

In eqs.~\ref{non-local dispersion1} the effect of non-locality is implicit in $\hat{G}_{NL}(k)$ and $\hat{G}_{NLD}(k,\Omega)$. A good  physical intuition can be established by studying two extreme cases for this effect. As demonstrated in Fig.~\ref{Fig3}, the interaction range decreases in thin cavities. In such cavities the interaction ultimately reduces to a local one with $G_{NL}(\vec{r};\vec{r}')\approx \delta(\vec{r} - \vec{r}')$, hence $\hat{G}_{NL}(k) = 1$, leading to a linear dispersion. Therefore, for a contact interaction the dispersion is separated to two distinct forms, one for free particles at large momenta ($\Omega \propto k^2$) and one for sonic modes  for small momenta ($\Omega \propto k$). 
The other extreme happens for thick cavities where the fluid becomes fully 3D with a long range, gravitational-type interaction where $G_{NL}(\vec{r};\vec{r}')=1/|\vec{r}-\vec{r}'|$. With $\hat{G}_{NL}(k)=1/k^2$, this Green's function would lead to a constant, $k$-independent dispersion at low momenta.

To provide a better understanding of the dependency of the dispersion in terms of physical parameters such as $L,N_{BEC}$, and $\alpha_{in}$, we define a critical momentum $k_{critical}$, as the largest momentum at which the dispersion is significantly modified by the interaction. Using the above given dispersion relations this parameter is determined as:

\begin{equation}
\label{critical momentum}
k_{critical} \le \frac{2\omega}{n_0}|\psi_{ss}(0)|\sqrt{2\frac{\alpha_{in} \beta c m_{ph}}{\kappa}\hat{G}(k_{critical})}~.
\end{equation}

Where $\hat{G}$ is a general representation for the Green's function and could be either of $\hat{G}_{NL}(k)$ for the non-local case, or $\hat{G}_{NLD}(\Omega,k)$ for the non-local and delayed one.

\begin{figure*}[h]
\centering
\includegraphics[scale=0.85]{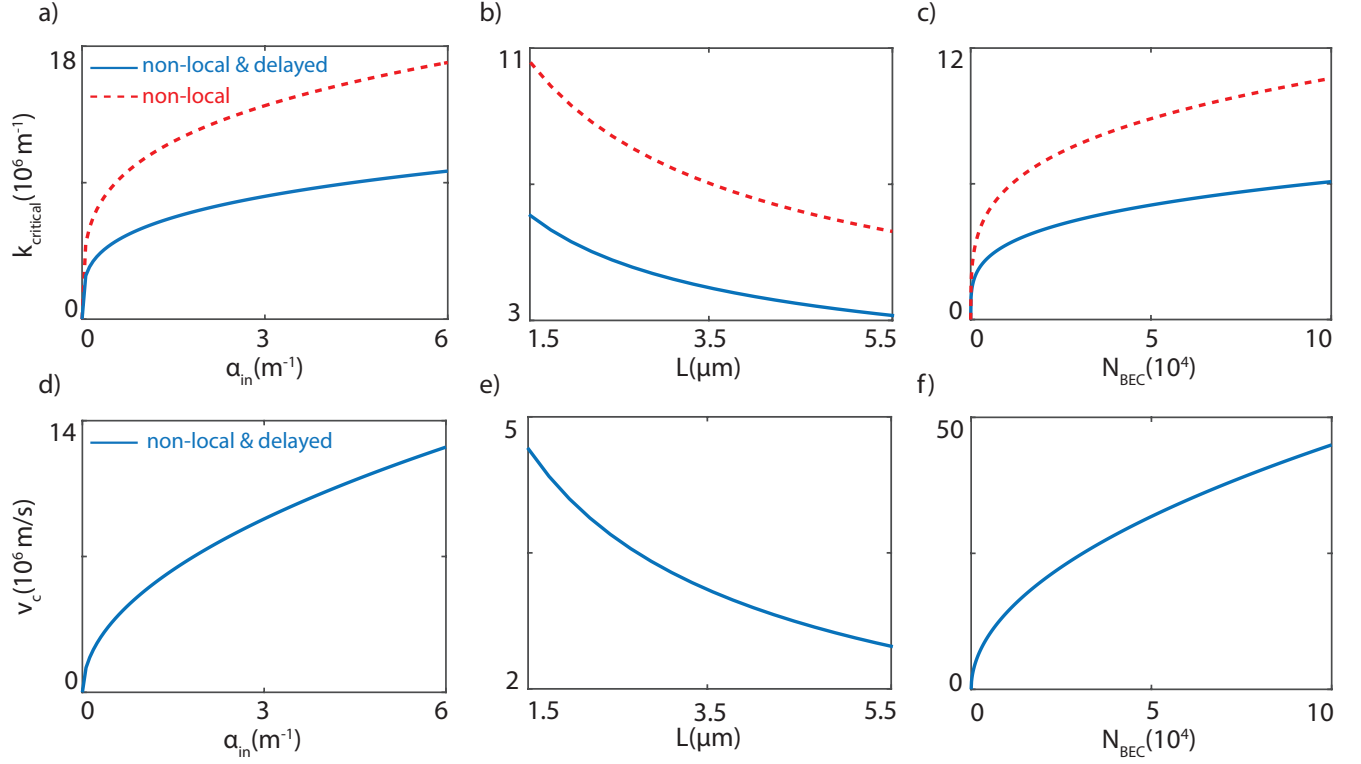}
\caption{\label{Fig5} The solid blue line gives the critical wavevector as a function of (a) the inelastic absorption $\alpha_{in}$ (for $L = 2~\mu m, N_{BEC}=6\times 10^4$), (b) the cavity length (for $\alpha_{in} =0.63~m^{-1}, N_{BEC}=6\times 10^4$), and (c) the number of photons in the condensate (for $\alpha_{in} = 0.63~ m^{-1} , L=2~\mu m$), accounting both for the delay and the non-locality of the thermo-optic effective photon interaction. For comparison, the dashed red lines give the critical wavevector when only the non-locality of this interaction, following eq.~\ref{non-local dispersion2}, is considered. The lower panels (d)-(f) show the behavoir of the critical velocity ($v_c$) for corresponding experimental parameters, where both the non-locality and the temporally delayed features of thermo-optic interaction are considered.
As described in the text, by increasing the interaction strength either via more absorption (a,d) or larger photon numbers (c,f), the interaction affects quasi-particles up to larger momenta, as seen from the larger critical wavevector as well as the larger critical velocity. Panels (b,e) show that the use of a larger cavity length, which makes the interaction more non-local, weakens the interaction effect, and reduces both the critical wavevector and the critical velocity.}
\end{figure*}

Figure~\ref{Fig5}(a)-(c) shows the behavior of the critical momentum as a function of some experimental parameters. The solid blue lines correspond to the predictions considering both delay and non-locality given by the Green's function of eq.~\ref{delayed, non-local dispersion}. For comparison, the dashed red lines are obtained when only considering the non-locality as given by the Green's function of eq.~\ref{non-local dispersion2}. While the critical wavevectors have similar dependencies to the experimental parameters, the inclusion of the temporal delay decreases the values. As discussed earlier, the inclusion of the temporal delay also results in complex eigenfrequencies $\Omega$ (Fig.~\ref{Fig4}(b)), leading to condensate instabilities.

The results of Fig.~\ref{Fig5}(a) and (c) show that the critical wavevector increases with both inelastic absorption coefficient $\alpha_{in}$ and photon number $N_{BEC}$ with corresponding dependency $\sqrt{\alpha_{in}}$ and $\sqrt{N_{BEC}}$, respectively. As shown in Fig.~\ref{Fig5}(b) the critical wavevector decreases for larger mirror separations $L$, when the interaction becomes more non-local.

According to Fig.~\ref{Fig4}(c), the thermo-optic interaction leads to a linear dispersion at low momenta, indicating the existence of sonic modes in this regime, which fulfills the Landau's criterion of superfluidiy. The slope of this line at low energies determines the velocity of sound in the condensate. Figure~\ref{Fig5}(d)-(f) shows the dependency of this critical velocity $v_c$ on the inelastic absorption coefficient, the cavity length, and the number of photons in the condensate, respectively, showing that the critical velocity increases with an increase in the inelastic absorption and photon number and decreases with increasing the cavity length.

\section{conclusions}
In this work we investigated the effect of a thermo-optic interaction, a temporally delayed and spatially non-local interaction, on a photon Bose-Einstein condensate in a dye-filled microcavity system. We derived the general form of the dispersion in such systems, calculated the spectrum of Bogoliubov modes in a plane-mirror microcavity, and identified a linear scaling for the low-energy modes. At larger transverse momenta, the usual quadratic free-particle dispersion of cavity photons is restored. The derived linear dispersion, corresponding to sonic modes, fulfills Landau's criterion for superfluidity. 

We envision several experimental and theoretical follow up studies based on the reported results here. To obtain accurate quantitative predictions, it would be important to investigate the implications of a finite heat conductivity of the mirrors on both the stationary and the dynamical features of the condensate subject to the thermo-optic non-linearity. 

For an experimental test of superfluidity based on the thermo-optic non-linearity a possible setup could use a photon fluid in a wedge-shaped cavity composed of two tilted flat mirrors, allowing the photon droplet to flow freely. By intentionally making a perturbation on one of the mirrors in the flowing path of the photon fluid, scattering would be observed if the fluid is dissipative. However, in the superfluid phase the photon droplet passes the perturbation without being scattered. By directly monitoring the photon condensate-defect interaction one should be able to distinguish between these two different phases.

Another follow up work could search for long-lived vortices of the photon condensate in a curved mirror microcavity. Along these lines, the implications of such a trapping potential on the Bogoliubov modes and the spectrum of elementary excitations should be studied. 

Another fascinating topic for future investigations is the search for analogies between the long-range thermo-optic interaction and gravitational physics and its consequences, including possible black-hole physics .

\section*{acknowledgment}
We thank Julian Schmitt, Sebastian Diehl, Iacopo Carusotto, Axel Pelster, and Jan Klaers for insightful discussions. H. A. acknowledges financial support from the Alexander von Humboldt foundation in terms of a postdoctoral fellowship. Partial financial support of the research from the DFG (CRC 185) and the ERC (INPEC) is appreciated as well.

\bibliography{ref}
\newpage

\end{document}